\documentclass[twocolumn,showpacs,preprintnumbers,amsmath,amssymb]{revtex4}

\usepackage{amsmath,amssymb}
\usepackage{graphicx}% Include figure files
\usepackage{dcolumn}% Align table columns on decimal point
\usepackage{bm}% bold math

\usepackage{graphicx,color}

\def\Beq{\begin{equation}}
\def\Eeq{\end{equation}}

\def\Bea{\begin{eqnarray}}
\def\Eea{\end{eqnarray}}

\def\Beaa{\begin{eqnarray*}}
\def\Eeaa{\end{eqnarray*}}

\def\BD{\begin{description}}
\def\ED{\end{description}}

\def\BC{\begin{center}}
\def\EC{\end{center}}

\def\Bcenter{\begin{center}}
\def\Ecenter{\end{center}}
\def\del{\partial}
\def\<{\langle}
\def\>{\rangle}
\def\({\left(}
\def\){\right)}

\def\leqa{\stackrel{<}{\sim}}
\def\geqa{\stackrel{>}{\sim}}

\def\a{\alpha}
\def\b{\beta}

\def\d{\delta}
\def\D{\Delta}

\def\l{\lambda}

\def\n{\nu}

\def\t{\tau}
\def\s{\sigma}
\def\^{\wedge}

\input{epsf.tex}

\begin{document}

\title{Smearing effect due to the spread of a probe-particle\\
 on the Brownian motion near a perfectly reflecting boundary}% Force line breaks with \\

\author{Masafumi Seriu}
 \email{mseriu@edu00.f-edu.u-fukui.ac.jp}
 \affiliation{%
 Department of Physics, Graduate School of Engineering, 
University of Fukui, 
Fukui 910-8507, Japan\\}%

\author{Chun-Hsien Wu}%
 \email{chunwu@phys.sinica.edu.tw}
\affiliation{%
Institute of Physics, 
Academia Sinica, Nankang, Taipei 11529, 
Taiwan 
}%


%%%%%%%%%%%%%%%%%%%%%%%%%%%%%%%%%%%%%%%%%%%%%%%%%%%%%%%%%%%%%%%%%
%%%%%%%%%%%%%%%%%%%%%%%%%%%%%%%%%%%%%%%%%%%%%%%%%%%%%%%%%%%%%%%%%
%ABSTRACT
%%%%%%%%%%%%%%%%%%%%%%%%%%%%%%%%%%%%%%%%%%%%%%%%%%%%%%%%%%%%%%%%%
%%%%%%%%%%%%%%%%%%%%%%%%%%%%%%%%%%%%%%%%%%%%%%%%%%%%%%%%%%%%%%%%%

\begin{abstract}
Quantum fluctuations of electromagnetic vacuum are investigated 
in a half-space bounded by  a perfectly reflecting  plate by introducing  a probe described by
 a charged wave-packet distribution in time-direction. 
The wave-packet distribution of the probe enables one to investigate 
 the smearing effect upon the measured vacuum fluctuations caused 
 by  the quantum nature of the probe particle.

It is shown that  the wave-packet spread of the probe particle significantly influences 
the measured velocity dispersion of the probe. 
 In particular, the asymptotic late-time behavior of its $z$-component, 
$ \langle \D v_{z}^{2} \rangle$, for the wave-packet case is quite different from the test point-particle case ($z$ is the coordinate 
normal  to the plate). 
The result for the wave-packet is $ \langle \D v_{z}^{2} \rangle \sim 1/\t^2$ in the late time ($\t$ is the measuring time), 
in stead of the reported late-time behavior  $ \langle \D v_{z}^{2} \rangle \sim 1/z^2$ for a point-particle probe.

This result can be quite significant for further investigations on  the measurement of vacuum fluctuations.
\end{abstract}

\maketitle

%%%%%%%%%%%%%%%%%%%%%%%%%%%%%%%%%%%%%%%%%%%%%%%%%%%%%%%%%%%%%%%%%
%%%%%%%%%%%%%%%%%%%%%%%%%%%%%%%%%%%%%%%%%%%%%%%%%%%%%%%%%%%%%%%%%
%SECTION 1
%%%%%%%%%%%%%%%%%%%%%%%%%%%%%%%%%%%%%%%%%%%%%%%%%%%%%%%%%%%%%%%%%
%%%%%%%%%%%%%%%%%%%%%%%%%%%%%%%%%%%%%%%%%%%%%%%%%%%%%%%%%%%%%%%%%
\section{\label{sec:Introduction}Introduction}

Physics of  vacuum fluctuations is one of the important research fields 
waiting for deeper understanding. 
One of the intriguing properties of quantum vacuum is 
that, even though it is within the realm of  microscopic physics, the vacuum itself is essentially a global concept 
and  that the spectral profile of the vacuum sensitively  reflects  global  conditions 
of the corresponding field. 
These subtle effects, mostly hidden in the macroscopic states,  sometimes become detectable in the vacuum environment.
These facts are most impressively illustrated by, e.g., 
Casimir effect and quantum noises in gravitational-wave detectors. 

One way of investigating  vacuum fluctuations is to 
study  the quantum Brownian motion of a test  particle coupled to the quantum field corresponding to the vacuum in question. 
In particular the velocity dispersion of the particle is a convenient quantity to investigate.
Along this line,  we have investigated 
vacuum fluctuations of the electromagnetic field near a perfectly reflecting mirror-boundary in the preceding paper~\cite{Switching}.
Starting with  the  model discussed by Yu and Ford~\cite{YuFord}, 
we have investigated in detail the {\it switching effect} in the measurement process by constructing 
a convenient switching function of time, consisting of a main measuring plateau with two Lorentzian tails.
The importance of studying the switching effect in the vacuum environment becomes  clear when one notes 
 that the switching process can be regarded as non-trivial, time-dependent interactions between 
the probe particle and the vacuum fluctuations. 
It has been  shown that the quantum vacuum fluctuations are, contrary to the usual macroscopic measurements, drastically influenced 
by the switching tails and that the anti-correlation between the main measuring part and the switching tails 
is playing an essential role in the process. 
 There the important factor has been the non-stational interactions between 
 the probe and the violent fluctuations in the vacuum, which is  basically a zero-sum environment. 
 
In the present paper, 
we go one step further to consider the {\it smearing effect} caused by the quantum nature of the probe particle upon, e.g., 
two-point correlation functions of the vacuum. 
In reality a  probe particle is  also a quantum object and should follow the quantum  principle. 
This fact becomes all the more important when the probed object is the quantum vacuum. 
Indeed,  as we have realized in our previous study~\cite{Switching}, 
various fluctuation spectra contained in the vacuum could show up in reality only through 
interactions with some other object (a ``probe" in a broad sense) so that the interplay between 
two time-scales of the vacuum-side  and the probe-side is an essential ingredient of  the process. 
Switching process is one place where these time-scales arise, which has given us a glimpse of a 
rich structure of the quantum vacuum. 
Now, the smearing effect due to the spread of the probe particle  is another place where 
the two scales get involved, and  its investigation  is expected to reveal some more aspects of the quantum vacuum. 

The most satisfactory treatment along this line may be to  quantize  the whole system (the field, the probe particle 
and the particles  composing the mirror) 
and see what happens.  However, clearly 
it is not  an easy task to  carry out   actual computations with such a treatment.

As the first step in this direction, then, we choose the most 
tractable way of  taking into account the quantum nature of 
a probe particle. In this paper, we describe  a probe particle  by a wave-packet. 
Here the  wave-packet nature of the probe particle should give rise to 
the spatial and temporal smearing  of the correlation functions.  
We here focus on the smearing in time-direction considering its affinity to our preceding work~\cite{Switching}.
Accordingly the term ``wave-packet"  below solely indicates a packet-like spread in the time-direction.

We  shall construct  a smeared two-point time-correlation functions by introducing  a Gaussian smearing function, 
and analyze their influence on the velocity dispersion of a probe. 
We shall find out that  the spread of 
the probe particle significantly influences the measured velocity dispersion, and 
the latter so obtained shows reasonable late-time behavior,  in the sense that 
it is compatible with other related situations such as the pure Lorentzian switching case~\cite{Switching}. 
 
Let us recall the original analysis in Ref.\cite{YuFord} to understand the importance of what these results 
imply. There, in a half-space bounded by a perfectly reflecting mirror, 
 the velocity dispersion of a test {\it point-particle} has been investigated in a ``sudden-switching" measuring process. Then, it has been  
 reported that the $z$-component of the  velocity dispersion, 
  $\< \D v_{z}^{2} \>$, does not  vanish in the late time, but shows an asymptotic late-time 
  behavior $\langle \D v_{z}^{2} \rangle \sim 1/z^2$ ($z$ is a distance from the mirror-boundary to the particle).  
  Ref.\cite{YuFord} interprets this  behavior as a transient effect due to a sudden switching process.  
What we shall find out below is that the late-time behavior of 
$\< \D v_{z}^{2} \>$ should be $\< \D v_{z}^{2} \> \sim 1/ \t^2$ ($\t$ is the  measuring time) rather than 
$\sim 1/z^2$,  {\it as far as 
the spread of the probe particle is taken into account}. 
Since it is more natural to regard the probe  a spreading particle (due to quantum effect) 
instead of a point-particle, one would see that the present result is more feasible.

In Sec.\ref{sec:formalism}, after setting up a model to be investigated, 
we prepare several useful formulas for analyzing the smearing effect on two-point
time-correlation functions. In Sec.\ref{sec:analysis}, we investigate 
the velocity dispersion of a probe-particle by applying formulas prepared in the preceding section. 
We first study the $z$-component $\< \D v_{z}^{2} \>$ and 
its late-time behavior. We also analyze in detail why the peculiar late-time behavior emerged 
in the case of the point-particle probe. Then the $x$- and $y$-components are also 
studied in a similar manner. 
Section \ref{sec:SummaryDiscussions} is devoted for a summary and several discussions. 
In {\it Appendix} \ref{app:1}, basic properties of the Lorentz-plateau switching function are 
summarized.

%%%%%%%%%%%%%%%%%%%%%%%%%%%%%%%%%%%%%%%%%%%%%%%%%%%%%%%%%%%%%%%%%
%%%%%%%%%%%%%%%%%%%%%%%%%%%%%%%%%%%%%%%%%%%%%%%%%%%%%%%%%%%%%%%%%
%SECTION 2
%%%%%%%%%%%%%%%%%%%%%%%%%%%%%%%%%%%%%%%%%%%%%%%%%%%%%%%%%%%%%%%%%
%%%%%%%%%%%%%%%%%%%%%%%%%%%%%%%%%%%%%%%%%%%%%%%%%%%%%%%%%%%%%%%%%
%%%%%%%%%%%%%%%%%%%%%%%%%%%%%%%
\section{\label{sec:formalism} Basic formulas for analysing smearing effect on  correlation functions}
%%%%%%%%%%%%%%%%%%%%%%%%%%%%%%%%

%%%%%%%%%%%%%%%%%%%%%%%%%%%%%%%%
\subsection{\label{sec:model} Mirror-boundary model for vacuum fluctuations}
%%%%%%%%%%%%%%%%%%%%%%%%%%%%%%%%

We start with the  model which has been 
 discussed in Ref.\cite{YuFord} and reanalyzed in Ref.\cite{Switching}.
 
 Preparing a flat, infinitely spreading mirror of perfect reflectivity 
 placed on the $xy$-plane  (i.e. $z=0$),  
 we consider the quantum vacuum of the electromagnetic field 
 inside  the half-space $z>0$.  
 To be more specific, we investigate the measurement process of  quantum fluctuations of the vacuum 
 probed by 
  a classical charged particle with mass $m$ and charge $e$. 
  When the velocity of the particle is much smaller than the light velocity $c$, 
  the particle couples only with the electric field $\vec{E} (\vec{x},t)$, and 
  the  motion for the particle is described by
%%%%%%%%%%%%%%%%%%%%%%%%%%%%%%%%%%%%%%%%%%%%%%%%%%%%%%%%%%%%%%%%%%%%%
\Beq
 m  \frac{d\vec{v}}{dt}= e \vec{E}(\vec{x},t)\ \ .
 \label{eq:eqofmotion}
\Eeq
%%%%%%%%%%%%%%%%%%%%%%%%%%%%%%%%%%%%%%%%%%%%%%%%%%%%%%%%%%%%%%%%%%%%%
If  the position of the particle does not change 
so much within the time-scale in question, Eq. (\ref{eq:eqofmotion}) along with 
the initial condition $\vec{v}(0)=\vec{v}_0$ is approximately solved as
%%%%%%%%%%%%%%%%%%%%%%%%%%%%%%%%%%%%%%%%%%%%%%%%%%%%%%%%%%%%%%%%%%%%%
\Beq
   \vec{v}(t) \simeq  \vec{v}_0 + \frac{e}{m}\  {\int_0}^t \vec{E}(\vec{x},t') dt'\ \ . 
 \label{eq:velocity}
\Eeq
%%%%%%%%%%%%%%%%%%%%%%%%%%%%%%%%%%%%%%%%%%%%%%%%%%%%%%%%%%%%%%%%%%%%%
Let us first consider the ``sudden-switching" case  investigated in Ref.\cite{YuFord}. 
Here the measurement is treated as a step-function-like process of 
abrupt switching on/off without no switching tails. It is characterized by a step-like 
switching function
%%%%%%%%%%%%%%%%%%%%%%%%%%%%%%%%%%%%%%%%%%%%%%%%%%%%%%%%%%%%%%%%%%%%%%%%%%
\Bea
\Theta (t) 
 &=& 1   \ \   ({\rm for}\ \ 0 \leq t \leq \t) \nonumber \\
 &=&  0  \ \  ({\rm otherwise})\ \ . 
 \label{eq:StepFun} 
\Eea
%%%%%%%%%%%%%%%%%%%%%%%%%%%%%%%%%%%%%%%%%%%%%%%%%%%%%%%%%%%%%%%%%%%%%%%%%%
In this case the velocity dispersions of the particle, $\< {\D v_i}^2 \>$ ($i=x, y, z$), are
given by 
%%%%%%%%%%%%%%%%%%%%%%%%%%%%%%%%%%%%%%%%%%%%%%%%%%%%%%%%%%%%%%%%%%%%%
\Beq
\< {\D v_i}^2 (\vec{x}, \t)\> = 
\frac{e^2}{m^2} \int_0^\t dt' \int_0^\t dt'' 
\< E_i (\vec{x} , t') E_i (\vec{x} , t'')  \>_R \ \ , 
\label{eq:v2sudden}
\Eeq
%%%%%%%%%%%%%%%%%%%%%%%%%%%%%%%%%%%%%%%%%%%%%%%%%%%%%%%%%%%%%%%%%%%%%
where $\< E_i (\vec{x} , t') E_i (\vec{x} , t'')  \>_R$ ($i=x, y, z$) are 
 the renormalized two-point correlation functions (the suffix ``R" is for ``renormalized").
 Here we note $\< E_i (\vec{x} , t)  \>_R=0$.  
Explicit expressions for $\< E_i (\vec{x} , t') E_i (\vec{x} , t'')  \>_R$ ($i=x, y, z$) are 
computed\cite{BroMac} as 
%%%%%%%%%%%%%%%%%%%%%%%%%%%%%%%%%%%%%%%%%%%%%%%%%%%%%%%%%%%%%%%%%%%%%
\Bea
 \< E_z (\vec{x} , t') E_z (\vec{x} , t'')  \>_R 
&=& \frac{1}{\pi^2} \frac{1}{ (T^2 - (2z)^2)^2} 
\label{eq:EzEz} \\
%%%%%%%%
 \< E_x (\vec{x} , t') E_x (\vec{x} , t'')  \>_R 
&=& \< E_y (\vec{x} , t') E_y (\vec{x} , t'')  \>_R \nonumber  \\
&& = - \frac{1}{\pi^2} \frac{T^2 + 4z^2}{ (T^2 - (2z)^2)^3} \ \ , 
\label{eq:ExEy}
\Eea
%%%%%%%%%%%%%%%%%%%%%%%%%%%%%%%%%%%%%%%%%%%%%%%%%%%%%%%%%%%%%%%%%%%%%
where   $T:= t'-t''$. 
(We set $c=\hbar=1$ hereafter throughout the paper.)

 Now in the previous paper\cite{Switching}, we have proceeded to a 
  generalization of  the above  model: 
We have considered the switching effect, introducing a more general  switching function rather than a simple 
step-like function Eq.(\ref{eq:StepFun}). 
In general, velocity dispersions  
are estimated by the integral  of  the form 
%%%%%%%%%%%%%%%%%%%%%%%%%%%%%%%%%%%%%%%%%%%%%%%%%%%%%%%%%%%%%%%%%%%%%%%%%%
\Beq
{\cal I} = \int_{-\infty}^\infty dt' \int_{-\infty}^\infty dt''\  F (t') F (t'') {\cal K} (t'-t'') \ \ ,  
\label{eq:intbasic}
\Eeq
%%%%%%%%%%%%%%%%%%%%%%%%%%%%%%%%%%%%%%%%%%%%%%%%%%%%%%%%%%%%%%%%%%%%%%%%%%
where the function $F (t)$ is an appropriate switching function mimicking 
an actual measurement process;  the integral kernel ${\cal K}$ is assumed to be an even  function of $T:=t'-t''$ with an appropriate asymptotic  behavior 
as $|T| \rightarrow  \infty$. In particular, 
when $F (t)$ is chosen to be the step-like function (Eq.(\ref{eq:StepFun})) 
and  $\cal K$ is chosen properly,   
Eq.(\ref{eq:intbasic}) reduces to the original formula Eq.(\ref{eq:v2sudden}).

%%%%%%%%%%%%%%%%%%%%%%%%%%%%%%%%%%%%%%%%%%%%%%%%%%%%%%%%%%%%%%%%%%%%%%%%%%%%%%%%%%
%%%%%%%%%%%%%%%%%%%%%%%%%%%%%%%%%%%%%%%%%%%%%%%%%%%%%%%%%%%%%%%%%%%%%%%%%%%%%%%%%%
\subsection{\label{sec:smearing} Smearing effect due to the spread of the probe particle}
%%%%%%%%%%%%%%%%%%%%%%%%%%%%%%%%%%%%%%%%%%%%%%%%%%%%%%%%%%%%%%%%%%%%%%%%%%%%%%%%%%%
%%%%%%%%%%%%%%%%%%%%%%%%%%%%%%%%%%%%%%%%%%%%%%%%%%%%%%%%%%%%%%%%%%%%%%%%%%%%%%%%%%%

The aim of the present paper is to proceed to  another direction of generalization of  the original  model; 
the {\it smearing effect} due to the quantum nature of the probe particle.
In reality, not only the electromagnetic field, but also the probe particle 
is  a quantum object and it cannot escape from the uncertainty principle. 
The quantum nature of the probe particle, thus, should cause the smearing effect on  the 
probed field. 
Treating the whole system as a fully quantum system 
is, however, neither practical nor easy for actual computations thought it might be most desirable. 
A more tractable way of  taking into account the quantum nature of 
the probe particle is, then,  to describe  the latter as a wave-packet. 
We here analyze  the smearing effect in time-direction 
induced by  the quantum nature of the probe particle, as a development of  the preceding investigations of 
the model~\cite{YuFord,Switching}. 
The term ``wave-packet"  in this paper is thus understood to indicate a packet-like spread in the time-direction.

Now the smearing effect is described by a suitable smearing function of the time variable $t$. Here 
we confine ourselves to  a Gaussian type function as a typical smearing function;
%%%%%%%%%%%%%%%%%%%%%%%%%%%%%%%%%%%%%%%%%%%%%%%%%%%%%%%%%%%%%%%%%%%%%
\Beq
   g(s-t)=\frac{1}{\sqrt{2\pi}b} \exp (-\frac{1}{2b^2}(s-t)^2) \ \ . 
 \label{eq:Gaussian}
\Eeq
%%%%%%%%%%%%%%%%%%%%%%%%%%%%%%%%%%%%%%%%%%%%%%%%%%%%%%%%%%%%%%%%%%%%%
Here the function $g(s-t)$ describes the Gaussian distribution of $s$ around 
the peak $t$ with  width $b$.

%%%%%%%%%%%%%%%%%%%%%%%%%%%%%%%%%%%%%%%%%%%%%%%%%%%%%%%%%%%%%%%%%%%%%%%%%%
At this stage, let us introduce  several variables and parameters 
for later convenience.
 
 First, let $\t$ be a time-scale characterizing the main measuring process. 
 In most cases,  a switching function $F(t)$ naturally determines  $\t$ (e.g. 
 Eq.(\ref{eq:StepFun}) and Eq.(\ref{eq:Lplat}) in {\it Appendix} \ref{app:1}). Furthermore, we introduce
%%%%%%%%%%%%%%%%%%%%%%%%%%%%%%%%%%%%%%%%%%%%%%%%%%%%%%%%%%%%%%%%%%%%%%%%%%
\Bea
&& T:=t' - t''\ ,\ \tilde{T}:=t'+t'' \ \ ,  \nonumber \\
&& S:=s'-s''\ ,\  \tilde{S}:=s'+s'' \ \ , \nonumber \\
&& \n :=(S-T)/\t \  ,  \  \b:= \sqrt{2}\, b/\t \ \ , \nonumber \\
&& \xi:=T/\t\ ,\   \eta= \tilde{T}/\t\ , \nonumber \\
&& \chi:=x/\mu \  , \   \l:= \n/\mu \ \ .
\label{eq:parameters}
\Eea 
%%%%%%%%%%%%%%%%%%%%%%%%%%%%%%%%%%%%%%%%%%%%%%%%%%%%%%%%%%%%%%%%%%%%%%%%%%

It is now straightforward to show that  the smeared kernel becomes  
%%%%%%%%%%%%%%%%%%%%%%%%%%%%%%%%%%%%%%%%%%%%%%%%%%%%%%%%%%%%%%%%%%%%%
\Bea
   \hat{\cal K}(T,\b)&=&\int_{-\infty}^\infty ds' \int_{-\infty}^\infty ds'' \  g(s'-t')\, g(s''-t'') \, {\cal K}(S) 
                                                            \ \ \nonumber \\
   &=& {\cal G}_\n (\b) \left[{\cal K}(T+\t \n)\right]  \ \ , \ \     
 \label{eq:KernelSmear}
\Eea
%%%%%%%%%%%%%%%%%%%%%%%%%%%%%%%%%%%%%%%%%%%%%%%%%%%%%%%%%%%%%%%%%%%%%
where ``${\cal G}_\n (\b)$" in 
the last line indicates the Gaussian integral-transformation with respect to $\n$ 
with the root-mean-square $\b$; it is defined as 
%%%%%%%%%%%%%%%%%%%%%%%%%%%%%%%%%%%%%%%%%%%%%%%%%%%%%%%%%%%%%%%%%%%%%
\Bea
{\cal G}_\n (\b) \left[ f(\n) \right] 
:= \frac{1}{\sqrt{2\pi} \b} \int_{-\infty}^\infty\, d\n\, e^{-\frac{\n^2}{2\b^2}} \, f(\n) \ \  \ \     
 \label{eq:GaussSmear}
\Eea
%%%%%%%%%%%%%%%%%%%%%%%%%%%%%%%%%%%%%%%%%%%%%%%%%%%%%%%%%%%%%%%%%%%%%
for any function $f(\n)$.
We note that $\frac{1}{\sqrt{2\pi} \b} e^{-\frac{\n^2}{2\b^2}} \rightarrow \d(\n)$ as $\b \rightarrow 0$, 
so that 
%%%%%%%%%%%%%%%%%%%%%%%%%%%%%%%%%%%%%%%%%%%%%%%%%%%%%%%%%%%%%%%%%%%%%
\Bea
\lim_{\b \rightarrow 0}{\cal G}_\n (\b) \left[ f(\n) \right]  = f(0) \ \ .
 \label{eq:deltalimit}
\Eea
%%%%%%%%%%%%%%%%%%%%%%%%%%%%%%%%%%%%%%%%%%%%%%%%%%%%%%%%%%%%%%%%%%%%%
Thus 
$\hat{\cal K} (T, \b) \rightarrow {\cal K}(T)$ as $\b \rightarrow 0$, recovering the original kernel for the point-particle limit. 
In view of  Eq.(\ref{eq:intbasic}), then, what we need to investigate is 
%%%%%%%%%%%%%%%%%%%%%%%%%%%%%%%%%%%%%%%%%%%%%%%%%%%%%%%%%%%%%%%%%%%%%
\Bea
\hat{{\cal I}} = \int_{-\infty}^\infty dt' \int_{-\infty}^\infty dt''
\  F (t') F (t'') \hat{\cal K} (T, \b) \ \ .
\label{eq:IntegralSmear}
\Eea
%%%%%%%%%%%%%%%%%%%%%%%%%%%%%%%%%%%%%%%%%%%%%%%%%%%%%%%%%%%%%%%%%%%%%
Due to linearity of the integral-transformation ${\cal G}_\n (\b)$, one can also 
represent 
$\hat{{\cal I}}$ as 
%%%%%%%%%%%%%%%%%%%%%%%%%%%%%%%%%%%%%%%%%%%%%%%%%%%%%%%%%%%%%%%%%%%%%
\Bea
\hat{{\cal I}} = {\cal G}_\n (\b) \left[ {\cal I}(\n) \right] \ \ ,
\label{eq:GaussInt}
\Eea
%%%%%%%%%%%%%%%%%%%%%%%%%%%%%%%%%%%%%%%%%%%%%%%%%%%%%%%%%%%%%%%%%%%%%
where
%%%%%%%%%%%%%%%%%%%%%%%%%%%%%%%%%%%%%%%%%%%%%%%%%%%%%%%%%%%%%%%%%%%%%
\Beq
{\cal I}(\n)= \int_{-\infty}^\infty dt' \int_{-\infty}^\infty dt''\ 
F (t') F (t'') {\cal K} (T+\t \n) \ \ .
\label{eq:Int(v)}
\Eeq
%%%%%%%%%%%%%%%%%%%%%%%%%%%%%%%%%%%%%%%%%%%%%%%%%%%%%%%%%%%%%%%%%%%%%
Equations (\ref{eq:GaussInt}) along with (\ref{eq:Int(v)}) serve as  general formulas for Gaussian smearing 
needed in our analysis. 

Let us note that ${\cal I}(\n)$ is an even function of $\n$ as far as 
the kernel ${\cal K}(T)$ is 
an even-function of $T$, irrespective of the  form of $F(t)$.   
Note also that, as $\b \rightarrow 0$,  $\hat{\cal I}$ reduces to ${\cal I}(\n =0)$, the original 
integral without smearing (Eq.(\ref{eq:intbasic})). 

In terms of the dimension-free variables $\xi$, $\eta$ (Eq.(\ref{eq:parameters})), ${\cal I}(\n)$ 
can also be expressed as
%%%%%%%%%%%%%%%%%%%%%%%%%%%%%%%%%%%%%%%%%%%%%%%%%%%%%%%%%%%%%%%%%%%%%
\Bea
&& {\cal I}(\n)= \frac{\t^2}{2} 
\int_{-\infty}^\infty d\xi \int_{-\infty}^\infty d\eta \ \nonumber \\
&& \qquad F (\frac{\t}{2}(\xi + \eta)) F (\frac{\t}{2}(\eta - \xi)) {\cal K} (\t(\xi + \n)) .
\label{eq:Int(v)2}
\Eea
%%%%%%%%%%%%%%%%%%%%%%%%%%%%%%%%%%%%%%%%%%%%%%%%%%%%%%%%%%%%%%%%%%%%%

Comparing the formula Eq.(\ref{eq:Int(v)}) with the one for the non-smearing case 
(Eq.(21) of the previous paper~\cite{Switching})), we see that the smearing of the probe particle causes   
a shift $\n$ in the argument of the kernel $\cal K$. This shift $\n$  reflects 
the effective spread of reference points ($t'$ and $t''$ in Eq.(\ref{eq:KernelSmear})) 
caused by the quantum spread of the probe particle, with $
\n=0$ corresponding to 
the non-smearing case. 
Then the total quantity $\hat{\cal I}$ is given by adding ${\cal I}(\n)$'s for various 
$\n$'s with the Gaussian weight as 
Eq.(\ref{eq:GaussInt}).

let us recall that the present model possesses two new features compared to  the original model of Ref.~\cite{YuFord}. 
One is the {\it switching effect} as already analyzed  in the previous paper~\cite{Switching} and the other 
is the {\it smearing effect}. Here in this paper, we focus on  the smearing effect and set aside 
the phenomena regarding the switching effect. 

For definiteness, let us choose  $F(t)$ as a Lorentz-plateau function $F_{\t \mu}(t)$ introduced in Ref.\cite{Switching}.
(Basic facts regarding the Lorentz-plateau function are summarized in {\it Appendix} \ref{app:1}.) 
The switching function $F_{\t \mu}(t)$  consists of a main measuring plateau-part and  two Lorentzian tails.  
We then focus on the main portion of the integral 
Eq.(\ref{eq:IntegralSmear}) 
coming from the integral region in which 
both the reference points $t'$ and $t''$ 
reside within the main measuring plateau-part. 

From now on, thus,  we focus on the first term of Eq.(\ref{eq:I(v)3}) 
(the suffix ``M" hereafter indicates ``measuring part") ,
%%%%%%%%%%%%%%%%%%%%%%%%%%%%%%%%%%%%%%%%%%%%%%%%%%%%%%%%%%%%%%%%%%%%%%%%%%
\Bea
 \hat{{\cal I}}_M := {\cal G}_\n (\b) \left[ {\cal I}_M(\n) \right] \ \ 
\label{eq:GaussInt_M}
\Eea
%%%%%%%%%%%%%%%%%%%%%%%%%%%%%%%%%%%%%%%%%%%%%%%%%%%%%%%%%%%%%%%%%%%%%%%%%%
with
%%%%%%%%%%%%%%%%%%%%%%%%%%%%%%%%%%%%%%%%%%%%%%%%%%%%%%%%%%%%%%%%%%%%%%%%%%
\Bea
 {\cal I}_{M}(\n) 
         = 2\t^2 \int_0^1 d\xi \ (1-\xi ) \left\{ {\cal K} (\t (\xi + \n)) \right\}_{{\cal S}(\n)} .
\label{eq:Int(v)_M}
\Eea
%%%%%%%%%%%%%%%%%%%%%%%%%%%%%%%%%%%%%%%%%%%%%%%%%%%%%%%%%%%%%%%%%%%%%%%%%%
Here we have introduced a symmetrization symbol $\{ \  \}_{\cal S}$ defined  as 
%%%%%%%%%%%%%%%%%%%%%%%%%%%%%%%%%%%%%%%%%%%%%%%%%%%%%%%%%%%%%%%%%%%%%%%%
\Beq
\left\{ f(u)  \right\}_{{\cal S}(u)}:=\frac{1}{2}\left( f(u) + f(-u) \right) \ \ 
\label{eq:symmetry}
\Eeq
%%%%%%%%%%%%%%%%%%%%%%%%%%%%%%%%%%%%%%%%%%%%%%%%%%%%%%%%%%%%%%%%%%%%%%%
for any function $f(u)$.

Note that this integral $\hat{{\cal I}}_M$
 is  to be compared with 
the result of the original analysis of Ref.~\cite{YuFord}: The latter  investigates 
the ``sudden-switching" case,  so that  
a switching function  $F (t)$ in Eq.(\ref{eq:intbasic}) is replaced by 
the step-like function $\Theta (t) = F_{\t \mu=0} (t)$ (Eq.(\ref{eq:StepFun})), which 
behaves in the same manner as  the plateau-part of the Lorentz-plateau function.

%%%%%%%%%%%%%%%%%%%%%%%%%%%%%%%%%%%%%%%%%%%%%%%%%%%%%%%%%%%%%%%%%
%%%%%%%%%%%%%%%%%%%%%%%%%%%%%%%%%%%%%%%%%%%%%%%%%%%%%%%%%%%%%%%%%
%SECTION 3
%%%%%%%%%%%%%%%%%%%%%%%%%%%%%%%%%%%%%%%%%%%%%%%%%%%%%%%%%%%%%%%%%
%%%%%%%%%%%%%%%%%%%%%%%%%%%%%%%%%%%%%%%%%%%%%%%%%%%%%%%%%%%%%%%%%
%%%%%%%%%%%%%%%%%%%%%%%%%%%%%%%
\section{\label{sec:analysis} Smearing effect on  the velocity dispersions}
%%%%%%%%%%%%%%%%%%%%%%%%%%%%%%%%

Basic formulas in hand, we can now investigate 
the velocity dispersions  
 originating from the main measuring part,  $\<\Delta v_i^2 \>_M$ ($i=x, y, z$). 
 
%%%%%%%%%%%%%%%%%%%%%%%%%%%%%%%%%%%%%%%%%%%%%%%%%%%%%%%%%%%%%%%%%
%%%%%%%%%%%%%%%%%%%%%%%%%%%%%%%%%%%%%%%%%%%%%%%%%%%%%%%%%%%%%%%%%
%SUBSECTION 3-1
%%%%%%%%%%%%%%%%%%%%%%%%%%%%%%%%%%%%%%%%%%%%%%%%%%%%%%%%%%%%%%%%%
%%%%%%%%%%%%%%%%%%%%%%%%%%%%%%%%%%%%%%%%%%%%%%%%%%%%%%%%%%%%%%%%%
%%%%%%%%%%%%%%%%%%%%%%%%%%%%%%%
\subsection{\label{sec:smearing_z} Smearing effect on  $\<\Delta v_z^2 \>$}
%%%%%%%%%%%%%%%%%%%%%%%%%%%%%%%%
 
We start with the analysis of $\<\Delta v_z^2 \>_M$. 
Plugging Eq.(\ref{eq:EzEz}) into ${\cal K}$ in Eq.(\ref{eq:Int(v)_M}), then, 
the velocity dispersion in the $z$-direction is given by 
%%%%%%%%%%%%%%%%%%%%%%%%%%%%%%%%%%%%%%%%%%%%%%%%%%%%%%%%%%%%%%%%%%%%%%%%%%
\Bea
\<\Delta v_z^2 \>_M = {\cal G}_\n (\b) \left[\<\Delta v_z^2 \>_M (\n) \right] \ \ .
\label{eq:Vz^2}
\Eea
%%%%%%%%%%%%%%%%%%%%%%%%%%%%%%%%%%%%%%%%%%%%%%%%%%%%%%%%%%%%%%%%%%%%%%%%%%
Here
%%%%%%%%%%%%%%%%%%%%%%%%%%%%%%%%%%%%%%%%%%%%%%%%%%%%%%%%%%%%%%%%%%%%%%%%%%
\Bea
\<\Delta v_z^2 \>_M (\n) &=& \frac{2e^2}{\pi^2m^2 \t^2} \left\{
                      \int_0^1 d\xi \frac{1-\xi}{\left((\xi + \n)^2 -\s^2\right)^2}\right\}_{{\cal S}(\n)} \nonumber \\
                    &=& \frac{2e^2}{\pi^2m^2 \t^2} \left\{ {\bf I}(\n) \right\}_{{\cal S}(\n)} \ \ , 
\label{eq:Vz^2-v}
\Eea
%%%%%%%%%%%%%%%%%%%%%%%%%%%%%%%%%%%%%%%%%%%%%%%%%%%%%%%%%%%%%%%%%%%%%%%%%%
with 
%%%%%%%%%%%%%%%%%%%%%%%%%%%%%%%%%%%%%%%%%%%%%%%%%%%%%%%%%%%%%%%%%%%%%%%%%%
\Bea
{\bf I}(\n) := I(\n , 1) + \n I(\n , 0)\ \ ,
\label{eq:bfIdef}
\Eea
%%%%%%%%%%%%%%%%%%%%%%%%%%%%%%%%%%%%%%%%%%%%%%%%%%%%%%%%%%%%%%%%%%%%%%%%%%
where we have introduced a basic integral
%%%%%%%%%%%%%%%%%%%%%%%%%%%%%%%%%%%%%%%%%%%%%%%%%%%%%%%%%%%%%%%%%%%%%%%%%%
\Bea
I(\n , \a) := \int_\n^{\n+1} du\  \frac{1-\a u}{(u^2 - \s^2)^2}\ \ .
\label{eq:Idef}
\Eea
%%%%%%%%%%%%%%%%%%%%%%%%%%%%%%%%%%%%%%%%%%%%%%%%%%%%%%%%%%%%%%%%%%%%%%%%%%

Furthermore, by virtue of a suitable contour integral,  the integral $I(\n , \a)$ can be estimated as 
%%%%%%%%%%%%%%%%%%%%%%%%%%%%%%%%%%%%%%%%%%%%%%%%%%%%%%%%%%%%%%%%%%%%%%%%%%
\Bea
I(\n , \a)\  \dot{=}\  J(\n , \a) - J(\n+1, \a) \ \ , 
\label{eq:Ical}
\Eea
%%%%%%%%%%%%%%%%%%%%%%%%%%%%%%%%%%%%%%%%%%%%%%%%%%%%%%%%%%%%%%%%%%%%%%%%%%
with 
%%%%%%%%%%%%%%%%%%%%%%%%%%%%%%%%%%%%%%%%%%%%%%%%%%%%%%%%%%%%%%%%%%%%%%%%%%
\Bea
J(\n ,\a):= \Re \int_\n^{\n+i\infty} dz \frac{1-\a z}{(z^2 - \s^2)^2}\ \ .
\label{eq:Jdef}
\Eea
%%%%%%%%%%%%%%%%%%%%%%%%%%%%%%%%%%%%%%%%%%%%%%%%%%%%%%%%%%%%%%%%%%%%%%%%%%
Here, following the notation of Ref.~\cite{Switching},  we introduce particular equality symbols 
$``\ \dot{=}\ "$ and $\approx$ (e.g. in Eq.(\ref{eq:VzResult3})) to remind us  that  a suitable regularization method \cite{DaviesDavies} has been applied. 

Now it is straightforward to compute the integral $J(\n , \a)$, yielding
%%%%%%%%%%%%%%%%%%%%%%%%%%%%%%%%%%%%%%%%%%%%%%%%%%%%%%%%%%%%%%%%%%%%%%%%%%
\Bea
J(\n , \a):= \frac{\nu-\a \s^2}{2\s^2 (\n^2 - \s^2)} - \frac{sgn(\n)}{8\s^3} \ln \left(\frac{|\n|+\s}{|\n|-\s}  \right)^2  \ ,
\label{eq:Jresult}
\Eea
%%%%%%%%%%%%%%%%%%%%%%%%%%%%%%%%%%%%%%%%%%%%%%%%%%%%%%%%%%%%%%%%%%%%%%%%%%
where $sgn(\n)$ indicates the sign of $\n$. 

Thus we get 
%%%%%%%%%%%%%%%%%%%%%%%%%%%%%%%%%%%%%%%%%%%%%%%%%%%%%%%%%%%%%%%%%%%%%%%%%%
\Bea
&& 
\<\Delta v_z^2 \>_M = \frac{4e^2}{\pi^2m^2 \t^2 \s^2} \times \nonumber \\
&& \qquad \qquad   \times 
{\cal G}_\n (\b) \left[ {\cal Z}(\s, 1+\n)-{\cal Z}(\s, \n) \right] 
\label{eq:VzResult}
\Eea
%%%%%%%%%%%%%%%%%%%%%%%%%%%%%%%%%%%%%%%%%%%%%%%%%%%%%%%%%%%%%%%%%%%%%%%%%%
where 
%%%%%%%%%%%%%%%%%%%%%%%%%%%%%%%%%%%%%%%%%%%%%%%%%%%%%%%%%%%%%%%%%%%%%%%%%%
\Bea
{\cal Z} (\s , \n) &=& \frac{|\n|}{16 \s} \ln\left(\frac{|\n|+\s}{|\n|-\s} \right)^2  \ \ ,  
\label{eq:Z-func}
\Eea
%%%%%%%%%%%%%%%%%%%%%%%%%%%%%%%%%%%%%%%%%%%%%%%%%%%%%%%%%%%%%%%%%%%%%%%%%%
which is an even-function of $\s$. 

%%%%%%%%%%%%%%%%%%%%%%%%%%%%%%%%%%%%%%%%%%%%%%%%%%%%%%%%%%%%%%%%%
%%%%%%%%%%%%%%%%%%%%%%%%%%%%%%%%%%%%%%%%%%%%%%%%%%%%%%%%%%%%%%%%%
%SUBSECTION 3-2
%%%%%%%%%%%%%%%%%%%%%%%%%%%%%%%%%%%%%%%%%%%%%%%%%%%%%%%%%%%%%%%%%
%%%%%%%%%%%%%%%%%%%%%%%%%%%%%%%%%%%%%%%%%%%%%%%%%%%%%%%%%%%%%%%%%
%%%%%%%%%%%%%%%%%%%%%%%%%%%%%%%
\subsection{\label{sec:QuantumProbe} Essential role of the probe as a quantum object}
%%%%%%%%%%%%%%%%%%%%%%%%%%%%%%%%

Let us pay attention to the mathematical structure of the formula for $\<\Delta v_z^2 \>_M$ in 
 Eq.(\ref{eq:VzResult}).

We observe that the formal $\s$-expansion of the factor 
${\cal G}_\n (\b) \left[ {\cal Z}(\s, 1+\n)-{\cal Z}(\s, \n) \right]$ 
in Eq.(\ref{eq:VzResult}) 
does not 
contain a constant term but  starts with the $O(\s^2)$-term.

This is shown by  first noting  that 
$\lim_{\s \rightarrow 0} {\cal Z} (\s , \n) =\frac{1}{4}$ (applying 
de L'Hopital theorem).  Then 
%%%%%%%%%%%%%%%%%%%%%%%%%%%%%%%%%%%%%%%%%%%%%%%%%%%%%%%%%%%%%%%%%%%%%%%%%%
\Bea
\lim_{\s \rightarrow 0} \left\{{\cal Z} (\s , 1 + \n)- {\cal Z} (\s , \n)\right\} =0\ \ . 
\label{eq:LimitRelation}
\Eea
%%%%%%%%%%%%%%%%%%%%%%%%%%%%%%%%%%%%%%%%%%%%%%%%%%%%%%%%%%%%%%%%%%%%%%%%%%
Thus ${\cal Z} (\s , 1 + \n)- {\cal Z} (\s , \n)$ does not contain a constant term w.r.t $\s$. Noting also that 
${\cal Z} (\s , 1 + \n)- {\cal Z} (\s , \n)$ is an even-function of $\s$ (see Eq.(\ref{eq:Z-func})), then, 
its $\s$-expansion should start with  the $O(\s^2)$-term.

Thus we can set
%%%%%%%%%%%%%%%%%%%%%%%%%%%%%%%%%%%%%%%%%%%%%%%%%%%%%%%%%%%%%%%%%%%%%%%%%%
\Bea
&& {\cal Z} (\s , 1 + \n) - {\cal Z} (\s , \n) = {\cal A}(\n) \s^2 +O(\s^4) \ \  \nonumber \\
&& {\rm with}   \nonumber \\
&& {\cal A}(\n) :=\lim_{\s \rightarrow 0} \frac{1}{\s^2}({\cal Z} (\s , 1 + \n) - {\cal Z} (\s , \n))\ \ .
\label{eq:O(s^2)}
\Eea
%%%%%%%%%%%%%%%%%%%%%%%%%%%%%%%%%%%%%%%%%%%%%%%%%%%%%%%%%%%%%%%%%%%%%%%%%%

Then we reach
%%%%%%%%%%%%%%%%%%%%%%%%%%%%%%%%%%%%%%%%%%%%%%%%%%%%%%%%%%%%%%%%%%%%%%%%%%
\Bea
 \<\Delta v_z^2 \>_M/\frac{4e^2}{\pi^2m^2 \t^2 \s^2}  &=&  
{\cal G}_\n (\b) \left[ {\cal A}(\n) \s^2 +O(\s^4) \right]  \nonumber \\
&=& {\cal A}(\b) \s^2 +O(\s^4)\ \ , 
\label{eq:VzResult2}
\Eea
%%%%%%%%%%%%%%%%%%%%%%%%%%%%%%%%%%%%%%%%%%%%%%%%%%%%%%%%%%%%%%%%%%%%%%%%%%
where ${\cal A}(\b):={\cal G}_\n(\b)\left[{\cal A}(\n)\right]$. 

Thus we conclude 
%%%%%%%%%%%%%%%%%%%%%%%%%%%%%%%%%%%%%%%%%%%%%%%%%%%%%%%%%%%%%%%%%%%%%%%%%%
\Bea
\<\Delta v_z^2 \>_M \approx  \frac{4e^2 {\cal A}(\b)}{\pi^2m^2 \t^2} + O\left((z/\t)^2\right) \ \ .
\label{eq:VzResult3}
\Eea
%%%%%%%%%%%%%%%%%%%%%%%%%%%%%%%%%%%%%%%%%%%%%%%%%%%%%%%%%%%%%%%%%%%%%%%%%%
Looking at Eq.(\ref{eq:GaussSmear}), only the integral-region $|\n| \leqa \b$ is important while 
it is reasonable to assume  $\b \ll 1$. Within this effective integral-region of $\n$ $( \leqa \b \ll 1)$, thus, it holds 
 ${\cal Z} (\s , 1 + \n) - {\cal Z} (\s , \n)>0$, so that ${\cal A} (\n) >0$.

Let us now estimate the factor ${\cal A}(\b)$ in Eq.(\ref{eq:VzResult3}), which is the Gaussian transformation of ${\cal A}(\nu)$ 
given by Eq.(\ref{eq:O(s^2)}). 
We first investigate the $\n$-dependence of ${\cal A}(\nu)$ as an integrand of  the Gaussian transformation (Eq.(\ref{eq:GaussSmear})).
Noting that 
%%%%%%%%%%%%%%%%%%%%%%%%%%%%%%%%%%%%%%%%%%%%%%%%%%%%%%%%%%%%%%%%%%%%%%%%%%
\Bea
\frac{\del}{\del \nu} {\cal Z} (\s , \n) = \frac{1}{16\s} \ln\left( \frac{|\nu| + \s}{|\nu| - \s} \right)^2 
+\frac{|\nu|}{4\s (\s^2 -|\nu|^2)}\ \ , 
\label{eq:delZ}
\Eea
%%%%%%%%%%%%%%%%%%%%%%%%%%%%%%%%%%%%%%%%%%%%%%%%%%%%%%%%%%%%%%%%%%%%%%%%%%
we see that 
${\cal Z} (\s , \n)$ does not change in effect in the $\n$-direction 
within an effective integral region 
$|\n| \leqa \b \ll 1$  in Eq.(\ref{eq:GaussSmear}). 
On the other hand, the same analysis shows that, the change of ${\cal Z} (\s , 1+\n)$ 
in the $\n$-direction is not negligible within the integral region in question. 
Then the $\nu$-dependence of  ${\cal A}(\nu)$ comes dominantly from the $O(\s^2)$-term of 
${\cal Z} (\s , 1 + \n)$, which is read off from a formal 
expansion of  ${\cal Z} (\s , 1+\n)$ 
 in a power series of $\s$, 
%%%%%%%%%%%%%%%%%%%%%%%%%%%%%%%%%%%%%%%%%%%%%%%%%%%%%%%%%%%%%%%%%%%%%%%%%%
\Bea
&& {\cal Z} (\s , 1+\n) = \frac{1}{4} + \frac{1}{12} \frac{\s^2}{|1+\n|^2} + O(\s^4) \ \ .
\label{eq:TaylorExp(1+nu)}
\Eea
%%%%%%%%%%%%%%%%%%%%%%%%%%%%%%%%%%%%%%%%%%%%%%%%%%%%%%%%%%%%%%%%%%%%%%%%%%
Thus one can estimate  ${\cal A}(\n) \simeq \frac{1}{12}\frac{1}{|1+\n|^2} \sim \frac{1}{12} + O(\nu)$. 
Furthermore the $O(\nu)$ term, the dominant part of which  is odd in $\n$,  
is negligible in the Gaussian transformation. 
One  thus get the estimation  ${\cal A}(\b) \sim \frac{1}{12}$ as far as $\b \ll 1$.  

From Eq.(\ref{eq:VzResult3}), we thus reach  the following result. 
{\it Taking into account the quantum spread of the probe particle, 
 the observed velocity dispersion $\<\Delta v_z^2 \>_M$ behaves as $1/\t^2$ 
in the late-time limit.} This behavior is consistent with 
other cases, e.g. the case with the Lorentzian switching~\cite{Switching}: Following the notations 
of {\it Appendix} \ref{app:1}, the latter case 
corresponds to  the limiting case $\t_1 \rightarrow 0$ with a fixed $\t_2$, 
characterized by  a Lorentzian function as a switching function ($F_{\t \mu}(t)=F_{0 \infty}(t)$). 

We see that the smearing due to the quantum spread of the probe particle  plays an essential role in 
deriving the above result.

Based on  the present framework, we can now understand the origin of 
a peculiar behavior $\<\Delta v_z^2 \>_M \sim 1/z^2$  reported in the original analysis 
of Ref.\cite{YuFord}. 
There the probe particle has been assumed to be a point particle, which corresponds to 
setting $\b = 0$ from the outset. It corresponds to  
take the limit $\b \rightarrow 0$ in Eq.(\ref{eq:VzResult}) with a fixed $\s$. 
Noting  Eq.(\ref{eq:deltalimit}) along with ${\cal Z}(\s , 0) =0$, then, we see that 
%%%%%%%%%%%%%%%%%%%%%%%%%%%%%%%%%%%%%%%%%%%%%%%%%%%%%%%%%%%%%%%%%%%%%%%%%%
\Bea
 \lim_{\b \rightarrow 0} \<\Delta v_z^2 \>_M &=&  \frac{4e^2}{\pi^2m^2 \t^2 \s^2}  {\cal Z}(\s, 1)\ \ \nonumber \\
                                             &=&  \frac{e^2}{4 \pi^2m^2 \t^2 \s^3} \ln\left(\frac{1+\s}{1-\s} \right)^2 \ \ . 
\label{eq:pointlimit}
\Eea
%%%%%%%%%%%%%%%%%%%%%%%%%%%%%%%%%%%%%%%%%%%%%%%%%%%%%%%%%%%%%%%%%%%%%%%%%%
Going back to the original variables, this result reads
%%%%%%%%%%%%%%%%%%%%%%%%%%%%%%%%%%%%%%%%%%%%%%%%%%%%%%%%%%%%%%%%%%%%%%%%%%
\Bea
           \<\Delta v_z^2 \> &\ \dot{=}\ &  \frac{e^2 \t}{32 \pi^2 m^2 z^3} \ln\left(\frac{\t+2z}{\t-2z} \right)^2 \nonumber \\
                             &\approx&  \frac{e^2}{4 \pi^2 m^2 z^2} +O(\left({z}/{\t}\right)^2) \ \ , 
\label{eq:originalresult}
\Eea
%%%%%%%%%%%%%%%%%%%%%%%%%%%%%%%%%%%%%%%%%%%%%%%%%%%%%%%%%%%%%%%%%%%%%%%%%%
exactly recovering the  result shown in Ref. \cite{YuFord}.

The essential point here is that, when one assumes 
the probe to be a point-particle (corresponding to setting $\b=0$),  
the factor ${\cal Z}(\s, \n)$ in Eq.(\ref{eq:VzResult}) {\it does not exist from the very beginning}, but only 
the factor ${\cal Z}(\s, 1+\n)$ appears. 
Then the cancellation of the constant term $\frac{1}{4}$  
due to the combination of 
${\cal Z}(\s, 1+\n) - {\cal Z}(\s, \n)$ (Eq.(\ref{eq:LimitRelation})) 
does not occur and 
the constant term  coming from ${\cal Z}(\s, 1+\n)$ 
gives rise to the peculiar behavior of $\<\Delta v_z^2 \> \sim 1/z^2$ in the late-time limit. 
Considering Eq.(\ref{eq:deltalimit}), we see that this phenomenon corresponds to   a {\it  measure-zero} set ($\n=0$) in 
the whole infinite $\n$-integral region, so that {\it it does not arise once the spread of the probe particle 
is taken into account}. 

We also note that the combination 
${\cal Z}(\s, 1+\n) - {\cal Z}(\s, \n)$ reminds us of 
the {\it anti-correlation} effect found in our previous work \cite{Switching}. 
There the similar combination has arisen from the interplay  between the main measuring part and 
the switching tails.  In the present case, it is understood that the temporal separation  between the center-part  
and the Gaussian tails of the probe particle is causing the similar effect.
Thus we find in Eq. (\ref{eq:VzResult})   the  anti-correlation 
effect caused by the spread of the probe. (See Sec.\ref{sec:SummaryDiscussions} for more arguments.)

%%%%%%%%%%%%%%%%%%%%%%%%%%%%%%%%%%%%%%%%%%%%%%%%%%%%%%%%%%%%%%%%%
%%%%%%%%%%%%%%%%%%%%%%%%%%%%%%%%%%%%%%%%%%%%%%%%%%%%%%%%%%%%%%%%%
%SUBSECTION 3-3
%%%%%%%%%%%%%%%%%%%%%%%%%%%%%%%%%%%%%%%%%%%%%%%%%%%%%%%%%%%%%%%%%
%%%%%%%%%%%%%%%%%%%%%%%%%%%%%%%%%%%%%%%%%%%%%%%%%%%%%%%%%%%%%%%%%
%%%%%%%%%%%%%%%%%%%%%%%%%%%%%%%
\subsection{\label{sec:smearing_xy} Smearing effect on  $\<\Delta v_x^2 \>=\<\Delta v_y^2 \>$}
%%%%%%%%%%%%%%%%%%%%%%%%%%%%%%%%

We can also analyze 
$\<\Delta v_x^2 \>_M$($=\<\Delta v_y^2 \>_M$) in a similar manner. 
(It suffices to consider only the $x$-component.)

Plugging Eq.(\ref{eq:ExEy}) into ${\cal K}$ in Eq.(\ref{eq:Int(v)_M}), 
the velocity dispersion in the $x$-direction is given by 
%%%%%%%%%%%%%%%%%%%%%%%%%%%%%%%%%%%%%%%%%%%%%%%%%%%%%%%%%%%%%%%%%%%%%%%%%%
\Bea
\<\Delta v_x^2 \>_M = 
{\cal G}_\n (\b) \left[\<\Delta v_x^2 \>_M (\n) \right] \ \ , 
\label{eq:Vx^2}
\Eea
%%%%%%%%%%%%%%%%%%%%%%%%%%%%%%%%%%%%%%%%%%%%%%%%%%%%%%%%%%%%%%%%%%%%%%%%%%
where
%%%%%%%%%%%%%%%%%%%%%%%%%%%%%%%%%%%%%%%%%%%%%%%%%%%%%%%%%%%%%%%%%%%%%%%%%%
\Bea
&& \<\Delta v_x^2 \>_M (\n) \nonumber \\
&&\qquad = - \frac{2e^2}{\pi^2m^2 \t^4} \left\{
                      \int_0^1 d\xi \frac{(1-x)((\xi + \n)^2 +\s^2)}{\left((\xi+\n)^2 -\s^2\right)^3}\right\}_{{\cal S}(\n)} \nonumber \\
                    &&\qquad = -\frac{2e^2}{\pi^2m^2 \t^2} \left\{ \tilde{{\bf I}}(\n) \right\}_{{\cal S}(\n)} \ \ , 
\label{eq:Vx^2-v}
\Eea
%%%%%%%%%%%%%%%%%%%%%%%%%%%%%%%%%%%%%%%%%%%%%%%%%%%%%%%%%%%%%%%%%%%%%%%%%%
with 
%%%%%%%%%%%%%%%%%%%%%%%%%%%%%%%%%%%%%%%%%%%%%%%%%%%%%%%%%%%%%%%%%%%%%%%%%%
\Bea
\tilde{\bf I}(\n) := {\bf I}(\n)
     + 2 \s^2 \left( \tilde{I}(\n , 1) + \n \tilde{I}(\n , 0)   \right)\ \ .
\label{eq:tilde_bfIdef}
\Eea
%%%%%%%%%%%%%%%%%%%%%%%%%%%%%%%%%%%%%%%%%%%%%%%%%%%%%%%%%%%%%%%%%%%%%%%%%%
Here  ${\bf I}(\n)$ is given by Eq.(\ref{eq:bfIdef}) with Eq.(\ref{eq:Idef}); another  basic integral 
$\tilde{I}(\n , \s)$ is defined by  
%%%%%%%%%%%%%%%%%%%%%%%%%%%%%%%%%%%%%%%%%%%%%%%%%%%%%%%%%%%%%%%%%%%%%%%%%%
\Bea
\tilde{I}(\n , \a) := \int_\n^{\n+1} du\  \frac{1-\a u}{(u^2 - \s^2)^3}\ \ .
\label{eq:tilde_Idef}
\Eea
%%%%%%%%%%%%%%%%%%%%%%%%%%%%%%%%%%%%%%%%%%%%%%%%%%%%%%%%%%%%%%%%%%%%%%%%%%
By virtue of a suitable contour integral, one can estimate $\tilde{I}(\n , \a)$ as 
%%%%%%%%%%%%%%%%%%%%%%%%%%%%%%%%%%%%%%%%%%%%%%%%%%%%%%%%%%%%%%%%%%%%%%%%%%
\Bea
\tilde{I}(\n , \a)\  \dot{=} \  \tilde{J}(\n , \a) - \tilde{J}(\n + 1 , \a)  \ \ ,
\label{eq:tilde_Ical}
\Eea
%%%%%%%%%%%%%%%%%%%%%%%%%%%%%%%%%%%%%%%%%%%%%%%%%%%%%%%%%%%%%%%%%%%%%%%%%%
with 
%%%%%%%%%%%%%%%%%%%%%%%%%%%%%%%%%%%%%%%%%%%%%%%%%%%%%%%%%%%%%%%%%%%%%%%%%%
\Bea
\tilde{J}(\n ,\a):= \Re  \int_\n^{\n+i\infty} dz \frac{1-\a z}{(z^2 - \s^2)^3}\ \ .
\label{eq:tilde_Jdef}
\Eea
%%%%%%%%%%%%%%%%%%%%%%%%%%%%%%%%%%%%%%%%%%%%%%%%%%%%%%%%%%%%%%%%%%%%%%%%%%
One can compute  $\tilde{J}(\n ,\a)$ as 
%%%%%%%%%%%%%%%%%%%%%%%%%%%%%%%%%%%%%%%%%%%%%%%%%%%%%%%%%%%%%%%%%%%%%%%%%%
\Bea
&& \tilde{J}(\n , \a) = -\frac{\a}{4(\n^2 - \s^2)^2} + \frac{\n(5\s^2 - 3 \n^2)}{8\s^4(\n^2-\s^2)^2} \nonumber \\
&& \qquad \qquad \qquad + \frac{3\ sgn(\n)}{32\s^5} \ln \left(\frac{|\n|+\s}{|\n|-\s}  \right)^2  \ \ .
\label{eq:tilde_Jresult}
\Eea
%%%%%%%%%%%%%%%%%%%%%%%%%%%%%%%%%%%%%%%%%%%%%%%%%%%%%%%%%%%%%%%%%%%%%%%%%%
Then we get
%%%%%%%%%%%%%%%%%%%%%%%%%%%%%%%%%%%%%%%%%%%%%%%%%%%%%%%%%%%%%%%%%%%%%%%%%%
\Bea
&& 
\<\Delta v_x^2 \>_M = -\frac{2e^2}{\pi^2m^2 \t^2 \s^2} \times \nonumber \\
&& \qquad \qquad   \times 
{\cal G}_\n (\b) \left[ {\cal W}(\s, 1+\n)-{\cal W}(\s, \n) \right] 
\label{eq:VxyResult}
\Eea
%%%%%%%%%%%%%%%%%%%%%%%%%%%%%%%%%%%%%%%%%%%%%%%%%%%%%%%%%%%%%%%%%%%%%%%%%%
where 
%%%%%%%%%%%%%%%%%%%%%%%%%%%%%%%%%%%%%%%%%%%%%%%%%%%%%%%%%%%%%%%%%%%%%%%%%%
\Bea
{\cal W} (\s , \n) &=& \frac{\n^2}{4(\n^2-\s^2)}- {\cal Z}(\s , \n)  \ \ ,  
\label{eq:W-func}
\Eea
%%%%%%%%%%%%%%%%%%%%%%%%%%%%%%%%%%%%%%%%%%%%%%%%%%%%%%%%%%%%%%%%%%%%%%%%%%
which is an even-function of $\s$. Noting that 
$\lim_{\s \rightarrow 0} {\cal W} (\s , \n) =0$ along with 
the even-function property of $\cal W (\s , \n)$ w.r.t. $\s$, we note that 
the $\s$-expansion of 
$\cal W (\s , \n)$ starts with the $O(\s^2)$-term. 
Compared with ${\cal Z}(\s , \n)$ in the previous subsection, we note that there is no  constant term in 
the formal $\s$-expansion of 
$\cal W (\s , \n)$. This is the origin of the apparent difference of the 
late-time behavior of $\<\Delta v_z^2 \>$ ($\sim 1/z^2$) and the ones for 
$\<\Delta v_x^2 \>= \<\Delta v_x^2 \>$ ($\sim 1/\t^2$) reported in the 
original analysis in Ref.\cite{YuFord}. 

Now we can set
%%%%%%%%%%%%%%%%%%%%%%%%%%%%%%%%%%%%%%%%%%%%%%%%%%%%%%%%%%%%%%%%%%%%%%%%%%
\Bea
&& {\cal W} (\s , 1 + \n) - {\cal W} (\s , \n) = {\cal B}(\n) \s^2 +O(\s^4) \ \  \nonumber \\
&& {\rm with}   \nonumber \\
&& {\cal B}(\n) :=\lim_{\s \rightarrow 0} \frac{1}{\s^2}({\cal W} (\s , 1 + \n) - {\cal W} (\s , \n))\ \ .
\label{eq:O(s^2)_W}
\Eea
%%%%%%%%%%%%%%%%%%%%%%%%%%%%%%%%%%%%%%%%%%%%%%%%%%%%%%%%%%%%%%%%%%%%%%%%%%
Then we see
%%%%%%%%%%%%%%%%%%%%%%%%%%%%%%%%%%%%%%%%%%%%%%%%%%%%%%%%%%%%%%%%%%%%%%%%%%
\Bea
 \<\Delta v_x^2 \>_M/(-\frac{2e^2}{\pi^2m^2 \t^2 \s^2})  &=&  
{\cal G}_\n (\b) \left[ {\cal B}(\n) \s^2 +O(\s^4) \right]  \nonumber \\
&=& {\cal B}(\b) \s^2 +O(\s^4)\ \ 
\label{eq:VxyResult2}
\Eea
%%%%%%%%%%%%%%%%%%%%%%%%%%%%%%%%%%%%%%%%%%%%%%%%%%%%%%%%%%%%%%%%%%%%%%%%%%
where ${\cal B}(\b):={\cal G}_\n(\b)\left[{\cal B}(\n)\right]$. 

We thus obtain 
%%%%%%%%%%%%%%%%%%%%%%%%%%%%%%%%%%%%%%%%%%%%%%%%%%%%%%%%%%%%%%%%%%%%%%%%%%
\Bea
\<\Delta v_x^2 \>_M = \<\Delta v_y^2 \>_M
\approx  - \frac{2e^2 {\cal B}(\b)}{\pi^2m^2 \t^2} + O\left((z/\t)^2\right) \ \ .
\label{eq:VxyResult3}
\Eea
%%%%%%%%%%%%%%%%%%%%%%%%%%%%%%%%%%%%%%%%%%%%%%%%%%%%%%%%%%%%%%%%%%%%%%%%%%

By the same estimation as the one for ${\cal A} (\n)$ (see 
Sec.\ref{sec:QuantumProbe}), 
 we see that 
 ${\cal W} (\s , 1 + \n) - {\cal W} (\s , \n)>0$, so that ${\cal B} (\n) >0$ as far as $\b \ll 1$.   
Furthermore, by a similar argument as for ${\cal A}(\b)$, one can estimate 
${\cal B}(\b) \sim \frac{1}{6}$ as far as $\b \ll 1$.

For confirmation, let us check that  our result reduces to  the original one in the point-particle limit, $\b \rightarrow 0$.  
Noting Eq.(\ref{eq:deltalimit}), we see that 
%%%%%%%%%%%%%%%%%%%%%%%%%%%%%%%%%%%%%%%%%%%%%%%%%%%%%%%%%%%%%%%%%%%%%%%%%%
\Bea
&& \lim_{\b \rightarrow 0} \<\Delta v_x^2 \>_M =  -\frac{2e^2}{\pi^2m^2 \t^2 \s^2}  {\cal W}(\s, 1)\ \ \nonumber \\
&& \ \ =   -\frac{2e^2}{ \pi^2m^2 \t^2 \s^2}
               \left\{ \frac{1}{4(1-\s^2)}- \frac{1}{16\s}\ln\left(\frac{1+\s}{1-\s} \right)^2 
               \right\} \ \ . \nonumber \\
&& {}
\label{eq:pointlimit2}
\Eea
%%%%%%%%%%%%%%%%%%%%%%%%%%%%%%%%%%%%%%%%%%%%%%%%%%%%%%%%%%%%%%%%%%%%%%%%%%
This is equivalent to 
%%%%%%%%%%%%%%%%%%%%%%%%%%%%%%%%%%%%%%%%%%%%%%%%%%%%%%%%%%%%%%%%%%%%%%%%%%
\Bea
          && \<\Delta v_x^2 \>=\<\Delta v_y^2 \> \nonumber \\
            & &\ \dot{=}\   -\frac{e^2 }{\pi^2 m^2} 
            \left\{
               \frac{\t^2}{8z^2(\t^2-4z^2)}
               -\frac{\t}{64z^3} \ln\left(\frac{\t+2z}{\t-2z} \right)^2  
            \right\} \nonumber \\
           & & \ \  \approx  - \frac{e^2}{3 \pi^2 m^2 \t^2} +O(\left({z}/{\t}\right)^2) \ \ , 
\label{eq:originalresult2}
\Eea
%%%%%%%%%%%%%%%%%%%%%%%%%%%%%%%%%%%%%%%%%%%%%%%%%%%%%%%%%%%%%%%%%%%%%%%%%%
exactly recovering the  result reported in Ref. \cite{YuFord}.

%%%%%%%%%%%%%%%%%%%%%%%%%%%%%%%%%%%%%%%%%%%%%%%%%%%%%%%%%%%%%%%%%
%%%%%%%%%%%%%%%%%%%%%%%%%%%%%%%%%%%%%%%%%%%%%%%%%%%%%%%%%%%%%%%%%
%SECTION 4
%%%%%%%%%%%%%%%%%%%%%%%%%%%%%%%%%%%%%%%%%%%%%%%%%%%%%%%%%%%%%%%%%
%%%%%%%%%%%%%%%%%%%%%%%%%%%%%%%%%%%%%%%%%%%%%%%%%%%%%%%%%%%%%%%%%
%%%%%%%%%%%%%%%%%%%%%%%%%%%%%%%
\section{\label{sec:SummaryDiscussions} Summary and discussions}
%%%%%%%%%%%%%%%%%%%%%%%%%%%%%%%%

In this paper, we have investigated the influence of  the quantum spread of a probe particle 
upon the measured fluctuations of quantum vacuum. 
Setting  up a model of the electromagnetic vacuum in 
a half-space with an infinite,  flat mirror-boundary with perfect reflectivity, we have analyzed 
the measurement process of the vacuum fluctuations  through the velocity dispersion of 
 a probe   particle, treated as a Gaussian wave-packet. 
 We have obtained  a smeared time-correlation functions, by which 
the velocity dispersion of the probe particle is computed.

As a result, it has been shown  that  the spread of the probe particle significantly influences the measured velocity dispersion; 
in particular the $z$-component, $ \langle \D v_{z}^{2} \rangle$, asymptotically behaves as 
$ \langle \D v_{z}^{2} \rangle \sim 1/\t^2$ as $\t \rightarrow \infty$,
which is quite different from $ \langle \D v_{z}^{2} \rangle \sim 1/z^2$ for a {\it point-particle} probe model reported in the literature~\cite{YuFord}. 
We believe that it is more realistic to represent a quantum particle by 
 a wave-packet   rather than by  a point-particle.

It has also been revealed that  the point-particle case mathematically 
corresponds to a measure-zero subset of  the whole set representing a spread-probe case. 
Provided that the probe-particle  is treated as  spreading, then, the measure-zero effect does not appear in the final result due to cancellation. 
On the other hand, when one starts with a point-particle probe from the out set (as so in Ref.\cite{YuFord}), 
only the contribution which would otherwise be regarded as measure-zero shows up, yielding 
the peculiar late-time behavior of $\< \D v_{z}^{2} \> \sim 1/z^2$. 

It has also been noted  that a particular combination of subtraction 
 appears in 
the velocity-dispersion formulas 
(i.e. the factor ${\cal Z}(\s, 1+\n) - {\cal Z}(\s, \n)$ in Eq. (\ref{eq:VzResult}) and, similarly Eq.(\ref{eq:VxyResult}))
has some similarity to 
the {\it anti-correlation} factor found in our previous work \cite{Switching}, which could be ascribed to 
the parallelism between the wave-packet-type time-distribution and the switching time-function. 

Let us recall once again that two different  time-scales are involved in the process. 
One is  a very short time-scale, say $\t_{vac}$, characterizing  typical vacuum fluctuations,  while the other is 
a much longer time-scale, say $\t_{probe}$, characterizing  the size of the probe. Then the following 
picture helps us  to depict the essence underlying all the  results obtained here.
Let us imagine taking the time-average of  a rapidly oscillating function with a typical  period $\t_{vac}$ 
within the time-interval $\t_{probe}$. One expects that strong cancellations occur in 
the normal case $\t_{vac} \ll \t_{probe}$, while 
no effective cancellation occurs in the case $\t_{vac} \sim \t_{probe}$. 
The former case has some affinity to the spread probe case discussed in the present paper, while 
the latter case to the point-particle model investigated in Ref.\cite{YuFord}.

Finally it should be mentioned  that the measuring time $\t$ cannot be infinitely long. 
First, the particle should not move so much in $z$-direction  during the measurement process because 
it is one of our assumptions in the beginning. (Due to the  symmetry of the configuration, only the $z$-direction matters.) 
Thus $\sqrt{\langle \D v_{z}^{2} \rangle}\ \t < z$.  
For later time $\t > 2z$, 
this inequality along with Eq.(\ref{eq:VzResult3}) turns to  the condition $z \geqa  O(10^{-2})\cdot \l_C $. 
(Here $\l_C=1/m$ is the Compton wave-length of the particle and $e^2 \sim 1/137$ in the present unit-system.) 
This is obviously satisfied. For earlier time $\t < 2z$, one can make use of the universal short-time behavior \cite{Switching} 
$\langle \D v_{z}^{2} \rangle \sim \frac{e^2 \t^2 }{m^2 z^4} $ and one gets the estimation $\t \leqa \sqrt{\frac{z}{e\ \l_C}}\ z $, 
which is also easily satisfied. 
Second, the wave-packet should not spread so much during the measuring process, since 
it violates the treatment of the probe as  a particle. 
Since the size of the wave-packet, $\D L$, spreads in later time as $\D L \sim \frac{\l_C}{b} \cdot \t$ \cite{BohmQM}, 
the restriction $\D L < z$ yields $\t < \frac{b}{\l_C } \cdot z$ (we recall that $b$ is the root-mean-square 
of the Gaussian wave-packet, Eq.(\ref{eq:Gaussian})). Thus it is desirable to prepare 
a wave-packet greater in size of, say, 10 times of the Compton length of the particle to study the late-time asymptotic behavior.

The spectral profile of the vacuum  can show up 
only through interactions with some kind of a probe so that 
the choice of a suitable probe is very essential in the measurement of the vacuum fluctuations. 
What we have learned here is that the approach taking into account of the quantum spread of a probe 
is quite essential in the measurement process. Considering that 
most of the approximation methods for studying quantum vacuums
so far rely on  a point-particle framework, 
it might be desirable and fruitful to investigate related problems once again along the line studied here.

\section*{Acknowledgement}
We would like to thank L. H. Ford for helpful comments 
on the present research.

%%%%%%%%%%%%%%%%%%%%%%%%%%%%%%%%%%%%%%%%%%%%%%%%%%%%%%%%%%%%%%%%%
%%%%%%%%%%%%%%%%%%%%%%%%%%%%%%%%%%%%%%%%%%%%%%%%%%%%%%%%%%%%%%%%%
%APPENDIX
%%%%%%%%%%%%%%%%%%%%%%%%%%%%%%%%%%%%%%%%%%%%%%%%%%%%%%%%%%%%%%%%%
%%%%%%%%%%%%%%%%%%%%%%%%%%%%%%%%%%%%%%%%%%%%%%%%%%%%%%%%%%%%%%%%%
\appendix

%%%%%%%%%%%%%%%%%%%%%%%%%%%%%%%%%%%%%%%%%%%%%%%%%%%%%%%%%%%%%%%%%
%%%%%%%%%%%%%%%%%%%%%%%%%%%%%%%%%%%%%%%%%%%%%%%%%%%%%%%%%%%%%%%%%
%APPENDIX 1
\section{\label{app:1} Lorentz-plateau switching function}
%%%%%%%%%%%%%%%%%%%%%%%%%%%%%%%%%%%%%%%%%%%%%%%%%%%%%%%%%%%%%%%%%
%%%%%%%%%%%%%%%%%%%%%%%%%%%%%%%%%%%%%%%%%%%%%%%%%%%%%%%%%%%%%%%%%
Let us  consider 
a Lorentz-plateau function $F_{\t \mu} (t)$ introduced in Ref.\cite{Switching}. 
It is constructed by gluing together a plateau part  and two Lorentzian tails; it is   
defined by 
%%%%%%%%%%%%%%%%%%%%%%%%%%%%%%%%%%%%%%%%%%%%%%%%%%%%%%%%%%%%%%%%%%%%%%%%%%
\Bea
F_{\t \mu} (t) 
 &=& 1 \qquad \qquad  \qquad \qquad \ \ \   ({\rm for}\ \ |t| \leq \t/2) \nonumber \\
 &=&   \frac{\mu^2}{ (|t|/\t- 1/2)^2+ \mu^2}  \ \  ({\rm for}\ \ |t| > \t/2)\  
 \label{eq:Lplat} 
\Eea
%%%%%%%%%%%%%%%%%%%%%%%%%%%%%%%%%%%%%%%%%%%%%%%%%%%%%%%%%%%%%%%%%%%%%%%%%%
 where $\t$ and $\mu$ are positive parameters. 
 Its plateau part corresponds to the stable measuring period, while two  Lorentzian tails  
 corresponding to switching-tails. 
 The time-scale for  the  measuring-part is $\t_1:=\t$, while 
 the time-scale of the switching-tails is characterized by 
 $\t_2:=\pi \mu \t$. Indeed 
%%%%%%%%%%%%%%%%%%%%%%%%%%%%%%%%%%%%%%%%%%%%%%%%%%%%%%%%%%%%%%%%%%%%%%%%%%
\Bea
&& \int_{-\t/2}^{\t/2} F_{\t \mu} (t) dt = \t =\t_1 \ \ , \ \ \nonumber \\
&& 2 \int_{\t/2}^{\infty} F_{\t \mu} (t) dt = \pi \mu \t =:\t_2\ \ . 
\label{eq:tau1tau2}
\Eea
%%%%%%%%%%%%%%%%%%%%%%%%%%%%%%%%%%%%%%%%%%%%%%%%%%%%%%%%%%%%%%%%%%%%%%%%%%
Then the dimension-free parameter 
%%%%%%%%%%%%%%%%%%%%%%%%%%%%%%%%%%%%%%%%%%%%%%%%%%%%%%%%%%%%%%%%%%%%%%%%%%
\Beq
\mu= \frac{\t_2}{\pi \t_1} 
\label{eq:mu}
\Eeq
%%%%%%%%%%%%%%%%%%%%%%%%%%%%%%%%%%%%%%%%%%%%%%%%%%%%%%%%%%%%%%%%%%%%%%%%%%
is the switching-duration parameter,  
describing   the relative switching duration compared to the main measuring time-scale. 

%%%%%%%%%%%%%%%%%%%%%%%%%%%%%%%%%%%%%%%%%%%%%%%%%%%%%%%%%%%%%%%%%
%%%%%%%%%%%%%%%%%%%%%%%%%%%%%%%%%%%%%%%%%%%%%%%%%%%%%%%%%%%%%%%%%
%APPENDIX 2
%\section{\label{app:2} Basic integral formula with the Lorentz-plateau switching function}
%%%%%%%%%%%%%%%%%%%%%%%%%%%%%%%%%%%%%%%%%%%%%%%%%%%%%%%%%%%%%%%%%
%%%%%%%%%%%%%%%%%%%%%%%%%%%%%%%%%%%%%%%%%%%%%%%%%%%%%%%%%%%%%%%%%

Now choosing  the Lorentz-plateau function 
$F_{\t \mu}(t)$ as a switching function $F(t)$,    Eq.(\ref{eq:Int(v)2}) is computed as 
%%%%%%%%%%%%%%%%%%%%%%%%%%%%%%%%%%%%%%%%%%%%%%%%%%%%%%%%%%%%%%%%%%%%%%%%%%
\Bea
 && {\cal I}(\n) 
         = 2\t^2 \int_0^1 d\xi \ (1-\xi ) \left\{ {\cal K} (\t (\xi + \n))\right\}_{{\cal S}(\n)} \nonumber \\
        &&  +   4 \pi  \mu^2 \t^2 \int_0^\infty d\chi\ \frac{1}{\chi^2 + 4 } 
         \left\{ {\cal K}(\mu \t (\chi+\l)\right\}_{{\cal S}(\l)}  \nonumber \\
         &&    +  4 \mu^2 \t^2 \int_0^\infty d\chi\ \left[
             \left\{  
                 {\cal K} ( \mu \t (\chi +\l) ) \right\}_{{\cal S}(\l)} \right. \nonumber  \\
            && \qquad \qquad \qquad  \left.  -  \left\{ {\cal K} ( \mu \t  (\chi+{1}/{\mu} +\l) \right\}_{{\cal S}(\l)} \right] {\cal F} (\chi)  \ \   \nonumber \\
\label{eq:I(v)3}
\Eea
%%%%%%%%%%%%%%%%%%%%%%%%%%%%%%%%%%%%%%%%%%%%%%%%%%%%%%%%%%%%%%%%%%%%%%%%%%
with
%%%%%%%%%%%%%%%%%%%%%%%%%%%%%%%%%%%%%%%%%%%%%%%%%%%%%%%%%%%%%%%%%%%%%%%%%%
\Beq
{\cal F}(\chi):= \left(1-   \frac{1}{\chi^2+4}  \right) \tan^{-1}\chi -  \frac{1}{\chi(\chi^2+4)}\ln (1+ \chi^2)\ \ .
\label{eq:F-function}
\Eeq
%%%%%%%%%%%%%%%%%%%%%%%%%%%%%%%%%%%%%%%%%%%%%%%%%%%%%%%%%%%%%%%%%%%%%%%%%%
Here, as is explained in the body of the paper (Eq.(\ref{eq:symmetry})), the symbol $\{ \  \}_{\cal S}$ 
indicates a symmetrization operator.

\vskip .5cm
%%%%%%%%%%%%%%%%%%%%%%%%%%%%%%%%%%%%%%%%%%%%%%%%%%%%%%%%%%%%%%%%%
%REFERENCES
%%%%%%%%%%%%%%%%%%%%%%%%%%%%%%%%%%%%%%%%%%%%%%%%%%%%%%%%%%%%%%%%%

\end{document}